\documentclass[12pt,a4paper]{article}
\usepackage{latexsym,epsf,epsfig}
\usepackage{graphicx}

\renewcommand{\d}{\partial}
\newcommand{\nn}{\nonumber\\}
\newcommand{\eqref}[1]{(\ref{#1})}
\newcommand{\exv}[1]{\left\langle{#1}\right\rangle}
\newcommand{\exvs}[1]{\langle{#1}\rangle}
\newcommand{\ph}{\varphi}
\newcommand{\rh}{\varrho}
\newcommand{\Tr}{\mathop{\textrm{Tr}}}
\newcommand{\pint}[2]{{\int\!\frac{d^{#1}#2}{(2\pi)^#1}\,}}

\renewcommand{\c}[1]{{\cal{#1}}}

\begin{document}

\title{Renormalization of O(N) model in 1/N expansion in auxiliary field
  formalism}

\author{
  A. Jakov\'ac\footnote{e-mail: jakovac@esr.phy.bme.hu}\\
  \scriptsize Department of Physics, University of Wuppertal, D-42097
  Wuppertal, Germany\footnote{On leave from Institute of Physics, BME
    Technical University, H-1521 Budapest, Hungary} } 

\date{August 13, 2008}

\maketitle

\begin{abstract}
  We study the renormalization of the O(N) model using the auxiliary field
  formalism (Hubbard-Stratonovich transformation) in the 1/N expansion at
  finite temperature. We provide the general strategy of renormalization for
  arbitrary order, and make calculation up to next-to-leading order. We show
  that renormalization is possible for any values of the condensates, prove
  the temperature independence of the counterterms and determine the cutoff
  dependence of the first nontrivial counterterm parts.
\end{abstract}

\section{Introduction}

Renormalization of approximations in quantum field theories which involve
resummation represents a complicated, much studied subject. Depending on the
system, various techniques are developed. Renormalization of mass and coupling
constant resummation was studied in \cite{BanMal},\cite{JSz}, in the framework
of 2PPI-resummation in \cite{Vers}. Renormalization of Hartree-Fock
resummation is performed in
\cite{Dest1},\cite{Dest2},\cite{PSz},\cite{FPSz}. The task of the
renormalization of 2PI resummation was solved by
\cite{HeesKnoll1},\cite{HeesKnoll2},\cite{HeesKnoll3},
\cite{BIR1},\cite{BIR2}, its generalization to n-point irreducible
resummations is discussed in \cite{BBRS1}. Application of this method to
scalar models \cite{AST},\cite{AR},\cite{BBRS2} and gauge theories
\cite{RS},\cite{BR} was done. The problem of how the 2PI counterterms can be
constructed is discussed in \cite{PSz2}. In \cite{AJ} a renormalization scheme
method was proposed to accomplish 2PI resummation and renormalization in a
single step. Renormalization of Schwinger-Dyson equations and its relation to
the 2PI approach, was discussed in \cite{CMD1},\cite{CMD2}.

Renormalization of O(N) model beyond the by-now textbook case leading order
\cite{Colemanetal} in $1/N$ expansion was studied recently in
\cite{ABW},\cite{AB}, using 2PI techniques in \cite{And}. In \cite{AB} the
authors presented a technique, with help of which they could provide a
renormalized, background dependent free energy up to next-to-leading
order. Although they used auxiliary field formalism (Hubbard-Stratonovich
transformation), after a point they switched to a simpler background,
eliminating the auxiliary field expectation value using its equation of motion
(EoM). It is not clear, whether there exists a consistent free energy also for
the auxiliary field background. The problem is that the auxiliary field
represents a composite operator of the original fields, and so it may show
unusual properties \cite{AB},\cite{BaRa}.

The primary goal of this paper is to propose a method to consistently
determine the auxiliary and scalar field dependent renormalized free
energy. It is important to define in a physically sensible way the pressure
(minus the saddle point value of the free energy density) and also the
different n-point functions at zero external momentum (from the derivatives of
the free energy near the saddle point). For a consistent renormalization one
has to know the (temperature and condensate-independent) values of the
counterterms in a given regularization. Here we use momentum cutoff
regularization, and we shall give the leading order values of the
counterterms.

The method proposed here represents an application of the general strategy
discussed in \cite{AJ}. We define a generalized renormalized Lagrangian, where
also the counterterms can be momentum dependent, but the bare theory remains
intact. The consistency of such renormalization schemes was studied in
\cite{AJ}, here we have to check if the asymptotic behavior of the propagator
fulfills the generic requirements as derived in \cite{AJ}.

The paper is organized as follows. First the definition of the model is
presented, then the renormalized Lagrangian is given in the most general
scheme which respects all symmetries of the Lagrangian. Then nontrivial
background fields are introduced. For renormalization we first fix the
renormalization conditions using the classical Lagrangian, then we determine
at leading and next-to leading order the values of the counterterms. We prove
that the counterterms are temperature independent, and give formulae for the
renormalized free energy. The paper is closed with conclusions.

\section{The renormalized Lagrangian}
\label{sec:renL}

The bare Lagrangian of the system reads:
\begin{equation}
  {\cal L} = \frac12(\d_\mu\bar\Phi_i)(\d^\mu\bar\Phi_i)
  -\frac{\bar m^2}2\bar\Phi_i\bar\Phi_i -\frac{\bar \lambda}{24 N}
  \left(\bar\Phi_i\bar\Phi_i\right)^2. 
\end{equation}
For tracing the $1/N$ powers, it is advantageous to split the quartic
interaction via the Hubbard-Stratonovich transformation. The above model
is equivalent to
\begin{equation}
  {\cal L} \equiv \frac12(\d_\mu\bar\Phi_i)(\d^\mu\bar\Phi_i) -\frac{\bar
    m^2}2\bar\Phi_i\bar\Phi_i - \frac12\bar\chi^2 -\frac{i \bar g}
  {2\sqrt{N}} \bar\chi \bar\Phi_i\bar\Phi_i,
\end{equation}
or, in the imaginary time
\begin{equation}
  \label{Lagbare}
  {\cal L}_E =  \frac12(\d_\mu\bar\Phi_i)(\d_\mu\bar\Phi_i) + \frac{\bar
    m^2}2\bar\Phi_i\bar\Phi_i + \frac12\bar\chi^2 + \frac{i \bar g}
  {2\sqrt{N}} \bar\chi \bar\Phi_i\bar\Phi_i.
\end{equation}
where
\begin{equation}
  \label{gandlambda}
  \bar g= \sqrt{\frac{\bar \lambda}3}.
\end{equation}

In order to be able to do perturbative computations, we should separate in the
Lagrangian the renormalized and the counterterm parts. We start with the wave
function renormalization:
\begin{equation}
  \bar\Phi= Z^{1/2} \Phi,\qquad \bar\chi = Z_\chi^{1/2}\chi_0.
\end{equation}
By introducing
\begin{equation}
  \label{baredef}
  m_0^2 = Z \bar m^2,\qquad g_0=ZZ_\chi^{1/2}\bar g,
\end{equation}
we can write
\begin{equation}
  \label{Lagwfr}
  {\cal L}_E = \frac Z2(\d_\mu\Phi_i)(\d_\mu\Phi_i) +\frac{m_0^2}2
  \Phi_i\Phi_i + \frac {Z_\chi}2 \chi_0^2 +\frac{i g_0}{2\sqrt{N}} \chi_0
  \Phi_i\Phi_i.
\end{equation}

As next we should renormalize the operators appearing in the Lagrangian. The
basic principle is that all operators which are scalar according to the
symmetry group of the Lagrangian and with positive mass dimensions must appear
in the renormalized Lagrangian. This means, that we have to include, in
addition to the operator set of \eqref{Lagwfr}, the operators $\chi_0$ and
$(\Phi_i\Phi_i)^2$ -- this latter is necessary, since in the actual form of
the Lagrangian \eqref{Lagwfr} the $(\Phi_i\Phi_i)^2$ and the $\chi_0^2$
operators are independent, and so they must be renormalized independently. It
may turn out, that we do not need counterterm for all these operators, then we
can omit them, but, for sake of completeness, we include all possibility
here. To allow later a full optimization of the propagator, we introduce a
generic kernel and quadratic counterterm for the $\chi_0$ fields. With the
notations
\begin{equation}
  Z=1+\delta Z,\qquad m_0^2=m^2+\delta m^2,\qquad Z_\chi = 1+\delta Z_\chi,
\end{equation}
the renormalized Lagrangian reads
\begin{eqnarray}
  {\cal L}_E && =  \frac12 \Phi_i \,(-\d^2+ m^2)\,\Phi_i + \frac12
  \chi_0\, H(i\d)\, \chi_0 + \frac{ig}{2\sqrt{N}} \chi_0\Phi_i\Phi_i + \nn
  &&+ \sqrt{N}\,i q\chi_0 +  \frac12 \Phi_i \,(-\delta Z \d^2 + \delta
  m^2)\,\Phi_i + \frac12 \chi_0\,\delta H(i\d)\, \chi_0 + \frac{i\delta g}
  {2\sqrt{N}} \chi_0\Phi_i\Phi_i + \frac{\delta \lambda}{24N}
  (\Phi_i\Phi_i)^2,
\end{eqnarray}
where the first row is the renormalized, the second is the counterterm
part. Some $N$-power factors are introduced just for convenience. The
consistency to the previous form of the Lagrangian requires (changing to
Fourier space for easier writing):
\begin{equation}
  \label{rest}
  H(p) + \delta H(p) = Z_\chi.
\end{equation}
The choice of the kernel is to large extent arbitrary, since -- if the above
constraint is fulfilled -- any choice leaves the bare Lagrangian, and so the
physics, untouched. That means that choices which depend on the environment
(temperature, singlet backgrounds, etc) are also possible. It is important to
note that, if $H(p)$ is chosen to be O(N) singlet, the Lagrangian remains O(N)
symmetric, and so in any perturbation theory based on this Lagrangian, the
consequences of the symmetry (ie. the Ward identities) remain valid.

A different question is whether a momentum-dependent counterterm can result
in, at all, a consistent perturbation theory. This question was addressed in
\cite{AJ}: if the kernel can be power expanded around asymptotic momenta, then
the Weinberg theorem is obeyed, so the renormalization program can be fully
implemented. We will check later that these conditions are really satisfied.

In this paper we aim to perform expansion with respect to $1/N$ to first
order (ie. up to ${\cal O}(1/N)$ in the free energy). As it will be later
confirmed, we assume the following dependence of the counterterms on
$1/N$:
\begin{eqnarray}
  \label{cntN}
  && q = q_0 + \sum_{n=1}^\infty \frac1{N^n} q_n,\qquad \delta
  f = \delta f_0 + \sum_{n=1}^\infty \frac1{N^n} \delta f_n,\nn&&
  \delta H = \delta H_0 + \sum_{n=1}^\infty \frac1{N^n} \delta H_n,\qquad
  \delta Z_\chi = \delta Z_{\chi,0} +  \sum_{n=1}^\infty \frac1{N^n} \delta
  Z_{\chi,n},\qquad\delta Z =  \sum_{n=1}^\infty \frac1{N^n} \delta Z_n,\nn&&
  \delta m^2 = \sum_{n=1}^\infty \frac1{N^n} \delta m_n^2,\qquad \delta g =
  \sum_{n=1}^\infty \frac1{N^n} \delta g_n,\qquad \delta \lambda = 
  \sum_{n=1}^\infty \frac1{N^n} \delta \lambda_n
\end{eqnarray}
where $\delta f$ is the counterterm for the free energy. Note that in certain
cases we have zeroth order counterterms as well!

In the perturbative calculation we fix the splitting of the bare parameters
into renormalized and counterterm pieces. This separation must be done in a
way that the divergences appearing in the expressions of physical quantities
cancel at each order. This determines the infinite part of the
counterterms. For the finite parts we require that certain physical quantities
have pre-defined values (renormalization scheme). For all divergent quantities
we should therefore find a physical quantity, which fixes the value of the
finite part; in our case we have seven potentially divergent quantities
($q,\delta f,\delta m^2, \delta Z,\delta Z_\chi,\delta g,\delta \lambda$), so
we need seven renormalization conditions. These will be defined through the
free energy and its derivatives later.

\section{Spontaneous symmetry breaking}

In the ground state of the system the bosonic scalar fields can
acquire expectation value. In general we can assume
\begin{equation}
  \exv{\chi_0} = -i\sqrt{N}\,X, \qquad \exv{\Phi_i} = \sqrt{N}
  \tilde\Phi_i.
\end{equation}
We will allow only those choices for $H(p)$, which depend on the O(N)
invariant combination of the background fields, ie. on
$\sum_i\tilde\Phi_i\tilde\Phi_i$. Then the complete Lagrangian will be
invariant under a simultaneous O(N) rotation of the background and the
fluctuation fields. Then, with an appropriate transformation we can
achieve that $\exv{\Phi_i} = 0$ for $i=1,\dots N-1$, while
\begin{equation}
  \exv{\Phi_N} = \sqrt{N} \Phi,\qquad \Phi^2 =
  \sum_i\tilde\Phi_i\tilde\Phi_i.
\end{equation}
As a consequence all results can depend only on
$\sum_i\tilde\Phi_i\tilde\Phi_i$, in particular the free energy, too. This
ensures that the first and second derivative of the free energy (effective
action) for the $k,j<N$ modes can be written as $f(X,\Phi_i)=V(X,\Phi^2)$, and
so
\begin{equation}
  \frac{\d f}{\d \tilde\Phi_i} = 2 \tilde\Phi_i\frac{\d V}{\d\Phi},\qquad
  \frac{\d^2 f}{\d \tilde\Phi_i\d\tilde\Phi_j} = 2\delta_{ij} \frac{\d
    V}{\d\Phi} + 4\tilde\Phi_j\tilde\Phi_k \frac{\d^2 V}{\d\Phi^2}.
\end{equation}
If at the saddle point $\d_\Phi V=0$ (we are in the broken phase), the
second derivative is a tensor with $N-1$ zero modes. Since the second
derivative of the free energy is the inverse propagator at zero momentum, we
find $N-1$ zero modes, independently on the value of $X$. So the Goldstone
theorem is satisfied, independently of $X$.

We introduce fluctuation fields, which already have zero expectation
value
\begin{equation}
  \chi_0 = -i\sqrt{N}\,X + \chi,\qquad \Phi_N = \sqrt{N} \Phi +\rh,
  \qquad \Phi_i=\ph_i\; (i=1\dots N-1).
\end{equation}
We write the resulting Lagrangian as
\begin{equation}
  {\cal L}_E = {\cal L}_E^{\mathrm{class}} +{\cal L}_E^{\mathrm{class, ct}} + \c
  L_E^{\mathrm{lin}} + {\cal L}_E^{(2)} + {\cal L}_E^{ct,I},
\end{equation}
where
\begin{eqnarray}
  && {\cal L}_E^{\mathrm{class}} = N \left[ \frac{m^2} 2 \Phi^2 -
    \frac{H(0)}2 X^2 + \frac{g}2 X \Phi^2 \right],\nn
  && {\cal L}_E^{\mathrm{class,ct}} = N \left[ \frac{\delta m^2} 2 \Phi^2 -
    \frac{\delta H(0)}2 X^2 + q X + \frac{\delta g}2
    X \Phi^2 + \frac{\delta \lambda}{24} \Phi^4 \right],\nn
  && {\cal L}_E^{\mathrm{lin}}= \sqrt{N}\left[ \left( m_0^2\Phi + g X\Phi + \delta
      g X\Phi +\frac{\delta\lambda}6 \Phi^3 \right) \rh + i\left( -X H(0)+
      \frac g2 \Phi^2 + q - X \delta H(0) + \frac{\delta g}2 \Phi^2
    \right)\chi \right],\nn  
  && {\cal L}_E^{(2)} =  \frac12 \chi H(i\d)\chi + \frac12 \ph_i
  \,\left(-\d^2+m^2 + gX \right)\,\ph_i +
  \frac12 \begin{array}[c]{cc} (\rh\;&\chi)\cr&\cr \end{array}
  \left(\begin{array}[c]{cc} -\d^2+m^2 + gX\quad & ig\Phi \cr 
      ig\Phi& H(i\d)\cr \end{array} \right)
  \left(\begin{array}[c]{c} \rh\cr \chi\cr \end{array}\right),\nn
  && {\cal L}_E^{ct,I}=  \frac{ig}{2\sqrt{N}} \chi (\ph_i\ph_i+\rh^2) +
  \frac12 \chi \delta H(i\d)\chi+ \frac12 \ph_i \,\left( -\delta
    Z\d^2+ \delta m^2 + \delta g X +\frac{\delta\lambda \Phi}6\right)
  \,\ph_i + \nn&&\qquad\qquad + \frac12 \begin{array}[c]{cc}
    (\rh\;&\chi)\cr&\cr \end{array} \left(\begin{array}[c]{cc} -\delta
      Z\d^2+\delta m^2 + \delta g X +\frac{\delta\lambda \Phi}2 \quad
      & i\delta g \Phi \cr i\delta g\Phi& \delta H(i\d)\cr \end{array}
  \right)\left(\begin{array}[c]{c} \rh\cr \chi\cr \end{array}\right)
  +\nn&&\qquad\qquad+ \frac{i\delta g}{2\sqrt{N}} \chi (\ph_i\ph_i
  +\rh^2) + \frac{\delta \lambda \Phi}{6\sqrt{N}}
  \rh(\ph_i\ph_i+\rh^2) + \frac{\delta \lambda}{24N}(\ph_i\ph_i+\rh^2)^2. 
\end{eqnarray}

The $\ph_i$ ($i=1\dots N-1$) modes will be called pions. Accordingly,
we will denote the tree level mass of the pions as:
\begin{equation}
  \label{pimass}
  m_\pi^2 = m^2+gX.
\end{equation}
Note that the so-defined pion mass depends on the condensate. The quadratic
term ${\cal L}_E^{(2)}$ provides the propagators, for which the following
notations will be used:
\begin{equation}
  G_\pi(p) = \frac1{p^2+m_\pi^2},\qquad \left(
    \begin{array}[c]{cc}
      G_{\rh\rh}\quad & G_{\rh\chi}\cr G_{\chi\rh}\quad& G_{\chi\chi} \cr
    \end{array}
    \right) = \left(\begin{array}[c]{cc} p^2+m_\pi^2\quad &
        ig\Phi \cr  ig\Phi& H(p)\cr \end{array} \right)^{-1},
\end{equation}
in particular
\begin{equation}
  G_{\chi\chi}(p) = \frac{p^2+m_\pi^2}{H(p)(p^2+m_\pi^2) + g^2\Phi^2}.
\end{equation}

We will compute the constrained free energy or effective potential, which is
the 1PI effective action for constant background; with the above separation we
can write
\begin{equation}
  f = {\cal L}_E^{\mathrm{class}} + \frac{N-1}2 \Tr \log\left(p^2+m_\pi^2\right)
  + \frac12 \Tr\log\det \left(\begin{array}[c]{cc} p^2+m_\pi^2\quad & ig\Phi
      \cr ig\Phi& H(p)\cr \end{array} \right) + \frac1{\beta V}
  \exv{1-e^{-S^{ct,I}}}_{1PI}.
\end{equation}
The first term provides the classical result, the rest is due to
quantum corrections. The interactions are treated through Taylor expansion:
\begin{equation}
  \frac1{\beta V} \exv{1-e^{-S^{ct,I}}}_{1PI} = \exv{{\cal L}^{ct,I}}_{1PI}
  -\frac12 \exv{\int {\cal L}^{ct,I} {\cal L}^{ct,I}}_{1PI}+\dots.
\end{equation}

\section{Renormalization}

After defining the generic framework we determine the necessary
counterterms. First we fix the physics by requiring some
renormalization conditions. Then order by order compute the necessary
diagrams, and require the cancellation of the divergences.

\subsection{Renormalization conditions}

To fix the renormalization conditions we will require that the radiative
corrections do not spoil the robust phenomena of the classical theory. To this
end we discuss the classical free energy. Without resummation it reads
\begin{equation}
  \frac{f_{cl}}N = \frac{m^2}2\Phi^2 - \frac12 X^2 + \frac g2 X\Phi^2.
\end{equation}
Its saddle point is at position
\begin{equation}
    \frac{\d f_{cl}}{\d X}\biggr|_{{\cal E}_{ref}} = -X_{\mathrm{min}} + \frac
  g2\Phi^2_{\mathrm{min}} = 0,\qquad 
  \frac{\d f_{cl}}{\d \Phi}\biggr|_{{\cal E}_{ref}} = \Phi_{\mathrm{min}}\left(
    m^2 + gX_{\mathrm{min}} \right) = 0.
\end{equation}
In the broken phase, where $m^2<0$ and $\Phi_{\mathrm{min}}\neq0$, its
solution reads
\begin{equation}
  X_{\mathrm{min}} =\frac{-m^2}g,\qquad \Phi_{\mathrm{min}}^2 =
  \frac{-2m^2}{g^2}.
\end{equation}
The first condition says that the pion mass at the saddle point
$m_\pi^2=m^2+gX_{\mathrm{min}} =0$, in accordance with the Goldstone theorem.
The rho-mass can be determined from the condition that the $\rh$-$\chi$
propagator at the saddle point of the free energy has a pole at
$p^2=-m_\rh^2$. Its position is found by requiring a zero for the determinant
of the $\rh$-$\chi$ kernel at the saddle point:
\begin{equation}
  \det\left(\begin{array}[c]{cc} -m_\rh^2 \quad &
        ig\Phi \cr  ig\Phi& H(-m_\rh^2) \cr \end{array} \right) = -m_\rh^2
    H(-m_\rh^2) + g^2\Phi^2 =0\qquad\Rightarrow\qquad
    m_\rh^2=g^2\Phi_{\mathrm{min}}^2 = -2m^2.
\end{equation}
Here we used that in the classical case $H(p)=1$.

In the symmetric phase, where $m^2$ is positive we obtain
$X_{\mathrm{min}}=\Phi_{\mathrm{min}}=0$, and $m_\pi^2=m_\rh^2=m^2$.

We will require later when we compute radiative corrections, that the position
of the saddle point, the value at the saddle point of the free energy and the
classical rho mass remain the same. We will also require that the residuum of
the pion propagator at $p^2=0$ is unity, there is no explicit $\Phi^4$ term,
and the coupling of the $X\Phi^2$ term is $g/2$. That is we will require the
following renormalization conditions:
\begin{eqnarray}
  \label{ren}
  &\displaystyle f_{{\cal E}_{ref}} =\frac{-X_{\mathrm{min}}^2}2,\qquad \frac{\d
    f}{\d X}\biggr|_{{\cal E}_{ref}}=0,\qquad \frac{\d f}{\d\Phi}\biggr|_{\c
    E_{ref}}=0, \qquad \frac{\d^3 f}{\d \Phi^2\d X}\biggr|_{\c
    E_{ref}}=g,\qquad \frac{\d^4 f}{\d \Phi^4}\biggr|_{{\cal E}_{ref}}=0,&\nn&
  \displaystyle \frac{\d G_\pi^{-1}(p)}{\d p^2}\biggr|_{{\cal E}_{ref}, p=0} =
  1,\qquad H(-m_\rh^2)=1,&
\end{eqnarray}
where we denoted 
\begin{equation}
  \label{ren_env}
  {\cal E}_{ref} = \{T=0, (X,\Phi)=(X,\Phi)_{\mathrm{min}}\}
\end{equation}
as the reference environment.

\subsection{Leading order}

Here we determine the ${\cal O}(N^0)$ counterterms, and give the leading order
free energy, as well as the leading order $\chi$ propagator.

The free $\chi$ propagator is ${\cal O}(N^0)$, but there are loop corrections
of the same order. Since these loop corrections are not suppressed by the
present series expansion parameter, $1/N$, they may appear in any diagram in
arbitrary number. That means that in the naive form at each level of
perturbation theory we have infinitely many diagrams. This phenomenon is
clearly must be avoided in a well defined perturbation theory, there the
radiative corrections must be of lower order in the expansion parameter than
the leading term. We can achieve this goal with help of our counterterm
$\delta H(p)$: we may tune it in a way that it cancels the ${\cal O}(N^0)$
radiative corrections to the $\chi$ propagator, or, in other words, we require
that the complete self-energy is zero for the $\chi$-field:
\begin{equation}
  \Sigma_{\chi\chi}(p,{\cal E})= -\frac{g^2}2 I(p,{\cal E}) -\delta H_0(p) =
  0 \qquad\Rightarrow\qquad \delta H_0(p)=-\frac{g^2}2 I(p,{\cal E}),
\end{equation}
where we explicitly signaled that the computation is performed in a generic
${\cal E} = \{ T, X, \Phi\}$ environment. The $I$ integral reads as
\begin{equation}
  I(p,{\cal E}) = \int_q G_\pi(p-q) G_\pi(q)\biggr|_{{\cal E}},
\end{equation}
it is the bubble diagram at environment ${\cal E}$. The integration symbol at
finite temperature corresponds to
\begin{equation}
  \int_q = T \sum_{n=-\infty}^\infty \pint3{\bf p},
\end{equation}
$p_0=2\pi n T$, and we apply an UV as well as an IR cutoff to regularize the
integrals. We can then calculate the result
\begin{equation}
  I(p,{\cal E}) = I_0(p) + I_T(p),
\end{equation}
where the zero temperature part reads
\begin{equation}
  I_0(p) = \frac1{16\pi^2} \left[ \ln \frac{e\Lambda^2}{\bar m_\pi^2} +
    \gamma \ln\frac{\gamma-1}{\gamma+1}\right],\qquad \gamma^2 =1+
  \frac{4\bar m_\pi^2}{p^2},
\end{equation}
where $\bar m_\pi^2 = m_\pi^2 + \Lambda_{IR}^2$ where $\Lambda_{IR}$ is the IR
cutoff. The finite temperature part computed with the Bose-Einstein
distribution function $n(\omega)$ is the following:
\begin{equation}
   I_T(p_0,p) = \frac1{8\pi^2 p} \int\limits_{m_\pi}^\infty d\omega
   \ln\left( \frac{p_0^2+(p\gamma + 2\omega)^2} {p_0^2+(p\gamma -
       2\omega)^2} \right) n(\omega).
\end{equation}

We may wonder that the counterterm we choose is environment
dependent. However, as we discussed in Section \ref{sec:renL}, the physics,
represented by the bare Lagrangian, is insensitive to the actual choice of the
counterterm, as far as the consistency equation \eqref{rest} is
fulfilled. Therefore we must choose:
\begin{equation}
  H(p) = 1 + \delta Z_{\chi,0} -\delta H_0(p) = 1+\delta
  Z_{\chi,0}+\frac{g^2}2 I(p,{\cal E})
\end{equation}
for the free $\chi$-kernel. Here $\delta Z_{\chi,0}$ is part of the bare
Lagrangian, and so it cannot be chosen environment dependent. But this is
actually unnecessary, since the bubble integral is only overall divergent, and
so the divergences do not depend on the IR quantities. Taking into account the
renormalization condition \eqref{ren} we shall choose
\begin{equation}
  \delta Z_{\chi,0} = -\frac{g^2}{32\pi^2} \ln\frac{e\Lambda^2}{m_\rh^2},
\end{equation}
then $H(p)$ is a finite quantity with $H(-m_\rh^2)=1$.

Some remarks in connection with the leading order result:
\begin{itemize}
\item With the procedure above we defined an optimally improved perturbation
  theory, where all net ${\cal O}(N^0)$ radiative corrections are missing (the
  different contributions cancel each other), and so, as a consequence, at
  each level only finite number of diagrams are generated. Therefore at higher
  orders there is no need to repeat this part of the procedure, only ordinary
  perturbation theory should be used.
\item One can realize that at asymptotic momenta the kernel $H(p)$ is
  logarithmic function of the momentum. This ensures that the
  renormalizability conditions of \cite{AJ} are satisfied.
\item Later we will need the ``double asymptotic'' form of $H$, when both the
  external $p$ as well as the internal loop momentum ($q$) go to infinity at
  the same rate. This form will be relevant for the analysis of overall
  divergences of higher order diagrams. The result is the same as the
  $m_\pi=0,\,T=0$ case, since these are IR quantities:
  \begin{equation}
    \label{Has}
    H_{as}(p) = 1 +\frac{g^2}{32\pi^2} \ln\frac{m_\rh^2}{p^2} =
    \frac{g^2}{32\pi^2} \ln\frac{L^2}{p^2},\qquad 
    L = e^{\frac{16\pi^2}{g^2}}m_\rh,
  \end{equation}
  $L$ is the position of the Landau pole.
\item The original bare coupling of the O(N) model is (cf. \eqref{gandlambda}
  and \eqref{baredef}) $\bar\lambda = 3g_0^2/(Z_\chi Z^2)$. In the leading
  order it reads
  \begin{equation}
    \bar\lambda = \frac{3g^2}{\displaystyle 1-\frac{g^2}{32\pi^2} \ln
      \frac{e\Lambda^2}{m_\rh^2}} = \frac{\lambda}{\displaystyle
      1-\frac{\lambda}{96\pi^2} \ln \frac{e\Lambda^2}{m_\rh^2}},
  \end{equation}
  where we defined $\lambda=3g^2$ as the renormalized quartic coupling. This
  is the ``nonperturbative renormalization'' formula for the coupling
  \cite{ABW}.
\end{itemize}

In order to be able to fix the other leading order counterterms, we shall
consider the free energy. The classical free energy is proportional to $N$,
but also the first quantum correction are of the same order of magnitude. The
interactions ($\exvs{1-\exp(-S^{ct,I})}$) do not contribute here, just the
free pion gas. In ${\cal L}^{\mathrm{class,ct}}$ we also have to take into
account those terms which contribute at leading order. What we have:
\begin{equation}
  \frac{f_0}N =\frac{m^2} 2 \Phi^2 - \frac{H(0)}2 X^2 + \frac g2 X \Phi^2
  + q_0X - \frac{\delta H_0(0)}2 X^2 +  \frac12 \Tr
  \log\left(-\d^2+m_\pi^2\right) + \delta f_0.
\end{equation}
We use that $H+\delta H=Z_\chi=1+\delta Z_\chi$, and we introduce an
environment independent UV regulator to write
\begin{equation}
  \frac{f_0}N =\frac{m^2} 2 \Phi^2 - \frac12 X^2 + \frac g2 X \Phi^2
  + q_0X - \frac{\delta Z_{\chi,0}}2 X^2 +  \frac12 \Tr
  \log\left(\frac{p^2+m_\pi^2}{p^2}\right) + \delta f_0.  
\end{equation}
The value of the $q_0$ and $\delta f_0$ counterterms can be determined by the
conditions \eqref{ren}. Taking into account the fact that our renormalization
conditions were fixed at zero temperature, we find:
\begin{eqnarray}
  q_0&&= -\frac g2 \pint4p G_\pi(p)|_{{\cal E}_{ref}}+\delta Z_{\chi,0}
  X_{\mathrm{min}} = \frac{g}{32\pi^2}\left[-\Lambda^2 + m^2
    \ln\frac{\Lambda^2}{m_\pi^2} + gX_{\mathrm{min}}
    \ln\frac{m_\rh^2}{em_\pi^2} \right],\nn
  \delta f_0&&=-\frac12\pint4p \log(\frac{p^2+m_\pi^2}{p^2}) -q_0
  X_{\mathrm{min}} +\frac{\delta Z_{\chi,0}}2X_{\mathrm{min}}^2=\nn&&=
  -\frac1{32\pi^2}\left[ m^2\Lambda^2 -\frac{m^4}2
    \ln\frac{\Lambda^2}{m_\pi^2} + \frac{g^2
      X_{\mathrm{min}}^2}2\ln\frac{m_\rh^2}{em_\pi^2} -\frac{m_\pi^4}4\right].
\end{eqnarray}
Note that the divergent parts of $q_0$ and $\delta f_0$ are
environment-independent, as it must be. As it can be easily checked, $\delta
Z_{\chi,0}$ cancels the divergence coming from the second derivative of the
free pion contribution. Then $f_0$ is a completely finite quantity.

\subsection{Next to leading order}

The next-to-leading order renormalization means that we must determine the
$1/N$ part of the counterterms. Due to the efforts we made at the leading
order, this order is the same as any other conventional perturbation theory:
since the problematic $\chi\chi$ self energy due to the pion modes is always
canceled by the $\delta H$ counterterm, we will have only a finite number of
diagrams to take into account. To determine the counterterms we should consider
physical quantities and require that the divergences cancel in the
perturbative expressions of these quantities.

First we determine the $\delta Z_1, \delta m^2_1, \delta g_1$ and $\delta
\lambda_1$ counterterms from the pion self-energy to the first order:
\begin{equation}
  \label{piself}
  \Sigma_{\pi}(p)= \frac{g^2}N \int_q G_\pi(p-q) G_{\chi\chi}(q) +\delta Z_1
  p^2 + \delta m_1^2 +\delta g_1 X + \frac{\delta\lambda_1} 6\Phi^2 +
  \frac{\delta\lambda_1} 6 \int_q G_\pi(q).
\end{equation}
The first term on the right hand side at first look is only overall divergent,
there are no divergent sub-diagrams, since it is only a one-loop diagram. In
the overall divergence all momenta go to infinity, and one could think that
the divergences cannot depend on the temperature, and the dependence on the
field background is polynomial: exactly of the form what are the second to
fifth terms in the above expression. But we must not forget, that here the
propagator is not the usual quadratic free propagator, and so its temperature
dependence can be stronger than an exponential suppression (in fact, in the
present case it is $H(p)\sim T^2/p^2$). At the same time the presence of the
last term is also embarrassing. It is clear that it is of order $1/N$:
although formally it is a next-to-next-to-leading order contribution, but the
number of pions ($N-1$) lifts up this contribution to the next-to-leading
order. But what is the role of this diagram?

\begin{figure}[htbp]
  \centering
  \begin{minipage}[b]{4cm}
    \includegraphics[width=4cm]{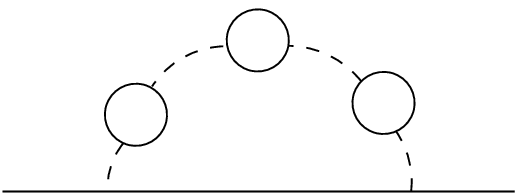}\\[0.7em]
    \centerline{a.)}
  \end{minipage}
  \hspace*{3em}
  \begin{minipage}[b]{4cm}
    \includegraphics[width=4cm]{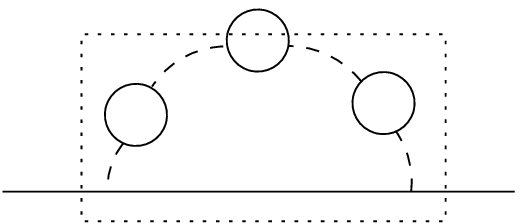}\\
    \centerline{b.)}
  \end{minipage}
  \hspace*{3em}
  \begin{minipage}[b]{2cm}
    \includegraphics[height=1.6cm]{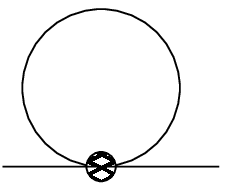}\\[0.3em]
    \centerline{c.)}
  \end{minipage}
  \caption{The pion self energy as expanded in the language of the
    non-resummed $\chi-\chi$ propagator (a.). Its potentially divergent
    subdiagram boxed by dotted line (b.) The diagram necessary to cancel this
    subdivergence (c.)}
  \label{fig:piself}
\end{figure}

If we rewrite the $\chi-\chi$ propagator in the original, not-resummed
language, the first contribution of \eqref{piself} corresponds to the diagram
of Fig.\ref{fig:piself}/a.
But this diagram \emph{has} a subdivergence, which is shown by Fig.1/b. To
cancel this subdivergence, we clearly want a counterterm diagram as appears in
Fig.1/c, which is exactly the same as the last term in the pion
self-energy. The conclusion is that in the original language \emph{the sum of
  the first and the last term} is overall divergent. This ensures that the
$\delta Z_1, \delta m_1^2, \delta g_1$ and $\delta \lambda_1$ counterterms are
temperature independent, which is needed for consistency. Their value can be
determined using the first term in \eqref{piself} where \emph{all momenta} in
the original, unresummed representation are asymptotic. In the resummed
language it means that we should use the asymptotic pion propagator, and the
``double asymptotic'' form of $H(p)$ described in \eqref{Has}.

To read off $\delta Z_1$ we should consider the $p^2$ dependence of the
overall divergent part of the first term in \eqref{piself} (the $\pi$-$\chi$
bubble integral). \eqref{Has} implies that the $\chi$-$\chi$ propagator
behaves as $G_{\chi\chi}(p)\sim 1/\ln(p^2)$. So we can write for the overall
divergent part of the bubble integral
\begin{equation}
  \int d^4q \frac1{(p-q)^2 \ln q^2} \sim \int^\Lambda \frac{dq\, q}{\ln q}
  \int\limits_0^\pi d\theta \frac{\sin^2\theta}{1 + (p^2/q^2) + 2p/q
    \cos\theta} \sim \int^\Lambda \frac{dq\, q}{\ln q} + \mathrm{finite}\times
  p^4.
\end{equation}
Therefore there is no divergent one-loop contribution proportional to
$p^2$. This implies
\begin{equation}
  \delta Z_1=0.
\end{equation}

For the determination of the cutoff dependence of $\delta m^2_1$ and $\delta
g_1$ we include also the pre-factors of \eqref{Has} and write the overall
divergent part of the bubble integral at zero external momentum
\begin{equation}
  \delta m^2_1 + \delta g_1 X + \frac4N\int\limits^{\Lambda^2}\frac{dq\,
    q^3}{q^2+m_\pi^2}\frac1{\ln L^2/q^2} =\mathrm{finite}.
\end{equation}
It means
\begin{equation}
  \delta m^2_{1,div} = \frac{\Lambda^2}{N \ln L/\Lambda}\left[1 - \frac1{2\ln
      L/\Lambda} + \frac1{2\ln^2L/\Lambda}\right] + \frac{2m^2}N \ln\ln
  \frac{L^2}{\Lambda^2},
\end{equation}
and
\begin{equation}
  \delta g_{1,div} = \frac{2g}N \ln\ln\frac{L^2}{\Lambda^2}.
\end{equation}
It is interesting to observe that neither $\delta m^2$ nor $\delta g$ is
suppressed by powers of the coupling, both are at the same order as $m^2$ and
$g$, respectively.

To have an expression for the $\delta\lambda_1$ it is simpler to consider
another observable, the 4-point function of the scalar fields. It can be
expressed at zero external momenta as
\begin{equation}
  \Gamma_{iijj,i\neq j}(0) = \frac{2g^4}{N^2} \int_q G_\pi(q)
  G_{\chi\chi}(q)G_\pi(q) G_{\chi\chi}(q) - \frac {\delta\lambda_1}N.
\end{equation}
At asymptotic momenta $G_\pi(q)\sim 1/q^2$ and $G_{\chi\chi}(q)\sim
1/\ln(q)$. Therefore the divergence structure is determined by
\begin{equation}
  \int \frac{dq\,q^3}{q^4(\ln q)^2} = \int \frac{d\ln q}{(\ln q)^2},
\end{equation}
which is also {UV finite} expression. That means that
\begin{equation}
  \delta\lambda_1=0.
\end{equation}
We should remark that the convergence of the one-loop contribution to the wave
function renormalization and to the four-point function does not mean that
these remain finite at any order. So we still have to keep $\delta Z_2$,
$\delta \lambda_2$ and all higher counterterms.

\begin{figure}[htbp]
  \label{fig:effpot}
  \centering
  \begin{minipage}[b]{2.5cm}
    \includegraphics[height=2.5cm]{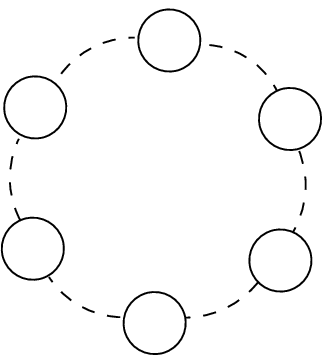}\\
    \centerline{a.)}
  \end{minipage}
  \hspace*{2em}
  \begin{minipage}[b]{2.5cm}
    \includegraphics[height=2.5cm]{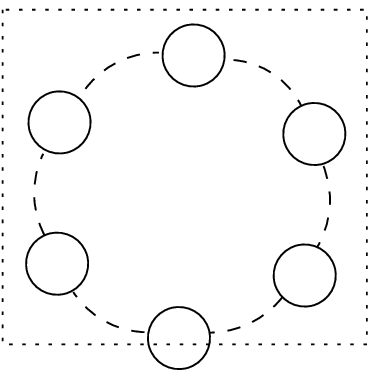}\\
    \centerline{b.)}
  \end{minipage}
  \hspace*{2em}
  \begin{minipage}[b]{1.5cm}
    \includegraphics[height=1.5cm]{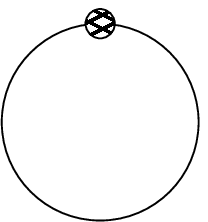}\\
    \centerline{c.)}
  \end{minipage}
  \hspace*{2em}
  \begin{minipage}[b]{2.5cm}
    \includegraphics[height=2.5cm]{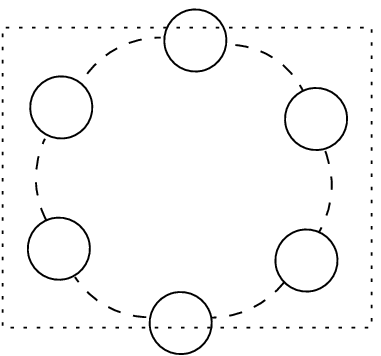}\\
    \centerline{d.)}
  \end{minipage}
  \hspace*{2em}
  \begin{minipage}[b]{1.5cm}
    \includegraphics[height=2.5cm]{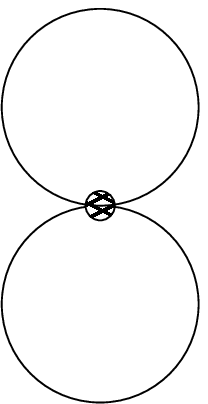}\\
    \centerline{e.)}
  \end{minipage}
  \caption{The trace-log contribution, expressed with help of the non-resummed
    propagators (a.). The quadratic subdivergence (b.) and its counterterm
    (c.). The quartic subdivergence (d.) and its counterterm (e.).}
\end{figure}

For the determination of the counterterms $q_1, \delta Z_{\chi,1}$ and $\delta
f_1$ we have to write up the effective potential to ${\cal O}(N^0)$. We have
\begin{eqnarray}
  \label{f1}
  f_1 =&&N \left[ \frac{\delta m_1^2} 2 \Phi^2 - \frac{Z_{\chi,1}}2 X^2 + q_1 X
    + \frac{\delta g_1}2 X \Phi^2 + \frac{\delta \lambda_1}{24} \Phi^4 \right]
   +\frac12 \Tr\log\biggl[H(p)(p^2+m_\pi^2) +g^2\Phi^2\biggr] +\nn&&+ \frac
   N2 \int_q G_\pi(q) \left[ \delta Z_1 p^2 + \delta m_1^2 +\delta g_1 X +
     \frac{\delta\lambda_1} 6\Phi^2\right] + \frac{N\delta\lambda_1}{24}
   \left[\int_q G_\pi(q)\right]^2 +\delta f_1.
\end{eqnarray}
This is the complete expression, which contains the cancellation of all
possible subdivergences. To understand the role of the different terms, the
logic is pretty much the same as in case of the pion propagator: the trace-log
contribution, although formally only overall divergent, if we expand it with
respect to the original, non-resummed modes, we find that it is, in fact, an
infinite sum of higher order diagrams, which contain also subdivergences. This
is illustrated in Fig.\ref{fig:effpot}.
The relevant diagrams necessary to cancel the subdivergences are the first and
second terms in the second line in \eqref{f1}, which are automatically
generated in the present framework. In the complete expression no
subdivergences survive, only the overall divergent or finite contributions
remain. Since in the overall divergent part all momenta go to infinity, there
is no way to generate a temperature dependent divergence. Therefore all the
necessary counterterms are also temperature independent.

The actual determination of the counterterms is analytically very hard, since
it will contain diagrams up to three loop order (each derivatives of $H(p)$
with respect to the background fields raise the diagrammatic loop number). On
the other hand if we use numerical techniques, the expression of $f_1$
contains only one-loop expressions. Since we \emph{know} that there exists a
consistent, temperature independent choice for $q_1, \delta Z_{\chi,1}$ and
$\delta f_1$, we can perform a zero temperature regularized numerical
calculation, do the derivations numerically, and determine the correct values
for the counterterms.

\section{Conclusions}

In this paper the renormalization of the O(N) model was discussed. Since the
radiative corrections can be of the same order of the expansion parameter (now
$1/N$) as the leading order result, a loop-order based perturbation theory is
possible only if we include the leading order quantum effects into the
quadratic Lagrangian. In 1PI technique it can be achieved by integrating out
the pionic modes \cite{AB}, in the 2PI case one applies an external propagator
field to represent the non-trivial propagation. In the present case we used a
non-conventional separation of the bare quadratic Lagrangian into free and
counterterm parts, where the free propagator contains all the leading order
corrections. In order to achieve this we need momentum dependent counterterms,
too. Since the so-defined free propagator satisfies the conditions of
\cite{AJ}, this perturbation theory can be treated with the normal
renormalization techniques. 

Once we have included the leading radiative corrections into the free
propagator, all the rest is conventional perturbation theory. We have to take
care, however, that the pion modes can still lift up certain diagrams to
higher level, but now it affects only a finite number of diagrams. If we
analyze the resulting perturbation theory in the language of the original
fields, we can see, that the unusual terms correspond to resummed
subdivergences of the leading order result. In our calculation these terms
appear automatically, one does not have to perform a detailed BPHZ
renormalization to obtain them.

As a result we established the cutoff dependence of the first nontrivial terms
of the counterterms, and gave a recipe how numerically can the rest be
computed. We gave also the expression for the renormalized free energy up to
next-to-leading order.

\section*{Acknowledgment}

The author acknowledges useful discussions with A. Patkos and Zs. Szep. The
work was supported by Hungarian Research Fund (OTKA) K68108. The author is
a recipient of a fellowship of the Humboldt foundation.

\end{document}